\documentclass[aps,pre,twocolumn,groupedaddress,showpacs,preprintnumbers,
amsmath,amssymb]{revtex4} 
\usepackage{graphicx} 
\usepackage{bm}
\begin{document}
\author{Joseph Rudnick} \author{Robijn Bruinsma}
\altaffiliation{Instituut-Lorentz/LION, Universiteit Leiden, 2300 RA,
Leiden, The Netherlands} 
\affiliation{Department of Physics and
Astronomy, UCLA, Box 951547, Los Angeles, CA 90095-1547} 
\date{\today}

\title{Icosahedral packing of RNA viral genomes}
\begin{abstract}
Recent studies reveal that certain viruses package a portion of their
genome in a manner that mirrors the {\em icosahedral symmetry} of the
protein container, or capsid.  Graph theoretical constraints forbid
exact realization of icosahedral symmetry.  This paper proposes a
model for the determination of quasi-icosahedral genome structures and
discusses the connection between genomic structure and viral assembly
kinetics.

\end{abstract}


\maketitle

An essential step in the assembly of any virus is the structural
reorganization of the single stranded (ss) or double stranded (ds) RNA
and DNA molecules that comprise the genome into a form that fits
inside the viral capsid.  This involves a well-known structural
incongruity; the protein capsid shells of nearly all sphere-like
viruses adopt \emph{icosahedral symmetry} while, as discussed below, a
viral genome of less than twelve segments fundamentally cannot assume
an icosahedral conformation in view of its one-dimensional, chain-like
primary structure \cite{bink}.  In fact, microscopy studies of the
structure of the ds DNA genome of bacteriophage viruses reveal a
completely \emph{non}-icosahedral spool-like organization
\cite{odijk}.  A semi-flexible chain---such as ds DNA---that is
confined inside a sphere having an interior surface that repels this
chain and a radius comparable or smaller than the chain persistence
length will, indeed, adopt such a spool-like structure as its free
energy minimum \cite{reimer,hud}.

The problem of the interior organization of ss RNA viral genomes is
more delicate.  SS nucleotide chains have a much greater
conformational flexibility than duplex chains, so ss RNA chains are
better able to adjust to the icosahedral symmetry of the capsid than
ds DNA chains.  The secondary structure of RNA molecules in solution
is characterized by paired segments conntected by branch points,
unpaired ``bubbles'' and ``hair pins'' (see Fig.  \ref{fig:fig2}). 
The various competing secondary structures can be analyzed by
statistical mechanical methods \cite{bundhwa}.  Moreover, unlike the
case of the bacteriophage viruses, for a significant number of ss RNA
viruses, capsid assembly \emph{requires} the presence of the genome
molecules \cite{fisher}.  This ``co-assembly,'' which can take place
spontaneously by self-assembly \cite{fisher}, is due in part to
non-specific electrostatic attraction between RNA molecules and capsid
proteins.  Attractive interactions between the genome and the interior
surface of the icosahedral capsid should promote icosahedral order. 
X-ray diffraction studies reveal that the outer layer of the genome of
a number of ss RNA viruses such as Cowpea Chlorotic Mottle Virus
(CCMV) \cite{bink,3d,ccmv}, Flock House Virus (FHV)
\cite{bink,3d,fhv}, calicivirus \cite{bink,3d,calicivirus}, Cowpea
Mosaic Virus (CMV) \cite{bink,3d,cmv}, Turnip Yellow Mosaic Virus
(TYMV) \cite{bink,3d,tymv}, indeed adopt at least partial icosahedral
symmetry.

\begin{figure}[htb]
\includegraphics[height=2in]{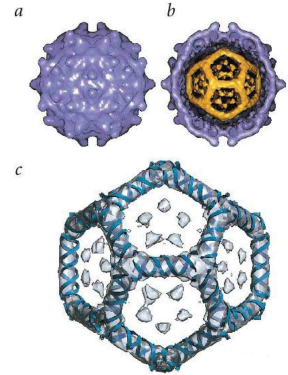} 
\caption{Image of the pariocoto virus, reconstructed from cryo-elctron
micrographs by Tang, et.  al.  \protect \cite{tang} }
\label{fig:fig1}
\end{figure}

\begin{figure}[htb]
\includegraphics[height=2in]{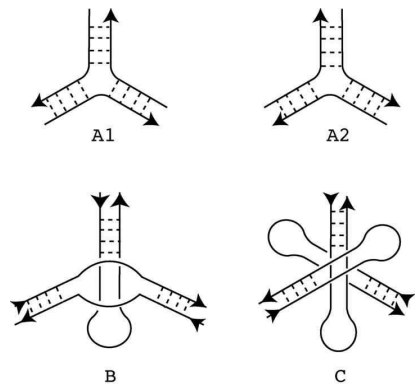}
\caption{The three types of vertices.  In the case of the ``A'' type
vertices (branch points) both a right-turn (A1) and a left-turn (A2)
vertex are shown. A type B vertex combines a ``bubble'' and a 
``hairpin,'' while a type C vertex contains only hairpins.}
\label{fig:fig2}
\end{figure}

A particularly striking illustration of icosahedral ordering is
provided by the \emph{Nodaviridae} group of viruses.  The FHV and
Pariacoto \emph{Nodaviridae} viruses have so-called ``T=3''
icosahedral capsids, as shown in Fig.  \ref{fig:fig1}.
The genome of these viruses consists of two single-stranded RNA
molecules.  One of these two molecules, RNA2, consists of about 1,400
(FHV), respectively 1,300 (Pariacoto) base pairs and encodes for
``Protein $\alpha$,'' a precursor of the capsid protein.  This
molecule plays a key stabilizing role in capsid assembly
\cite{fisher}, and it is probably the part of the genome that is
resolved in the diffraction experiments.  As shown in Figs. 
\ref{fig:fig1}b and c, the icosahedral portion of the RNA genome is
distributed over a \emph{dodecahedral cage} formed by the
low-curvature borders of twelve five-fold pyramids that together make
up the T=3 rhombic tricontahedron capsids of the FHV and Pariacoto
viruses.  Even though the genome is ss RNA, the portions shown in Fig. 
\ref{fig:fig1} actually have the form of \emph{double helices} with
the two strands oriented in the standard way with opposite 3' to 5'
directions.  These paired double-helical segments have a length per
edge of about 10 bases for the FHV structural study and 25 bases for
the Pariacoto structural study.  Given the 30 edges of a dodecahedron,
this means that for FHV about 43\% of RNA2 contributes to the
double-helical sections, and for Pariocoto \emph{essentially 100\% of
RNA2}.  If we assume that the 20 (unresolved) \emph{vertices} are also
structurally identical, we would have to connect together the 30
double-helical segments with 20 identical three-strand branch-points
(see Fig.  \ref{fig:fig2}A) placed at the vertices of the
dodecahedron.  The resulting structure has true icosahedral symmetry
(apart from the different base-pair sequences for the different
segments), {\em but it actually consists of 12 separate, interlocking
RNA rings}, which is inconsistent with the structure of the
\emph{Nodaviridae} genome \cite{tangnote}.  This structural conflict
is closely connected to a classical result of graph theory, which
states that it is impossible to construct a one-dimensional path
restricted to the edges of either an icosahedral or a dodecahedral
structure while visiting every edge just once
\cite{bollobas,graphnote} (a dodecahedron has the same symmetry
properties as an icosahedron).  The proposition that the 20 vertices
obey icosahedral symmetry is thus incorrect.  As shown below,
information concerning the structural inhomogeneity of the vertices
actually provides us with important information on the \emph{formation
history} of the virus.  It is the aim of the present article to
present a statistical mechanical model for quasi-icosahedral ordering
of ss RNA viral genomes.

On the basis of the above-mentioned X-ray structure studies, we 
set the following constraints on the  secondary structure of RNA2: 
\begin{enumerate}

\item[(i)] the edges of the dodecahedron should be occupied by the
rigid duplex segments.

\item[(ii)] the branch points, bubbles, and hair-pins of
the secondary structure should be confined to the vertex regions.

\end{enumerate}

Under these constraints, there are only three possible types of vertex
structures, which can be represented diagrammatically as follows.  The
first vertex type (``A'') are pure branch-points, the familiar
three-strand junctions of ds RNA and DNA. These vertex types are
illustrated in Fig.  \ref{fig:fig2}.  The arrows in the figure
indicate the 3' to 5' direction of the RNA2 molecule.  Electrostatic
repulsion between the duplex branches of the junction favors
$120^{\circ}$ angles between the branches, which have to be deformed
by a moderate amount in order to fit on the vertex of a dodecahedron. 
Cases A1 (a ``right-turn'' vertex) and A2 (a ``left-turn'' vertex) are
related by a rotation over 180 degrees along one of their branches. 
The energy cost, $\Delta E_{A}$, is the same for the two cases.

Next, a type ``B'' vertex consists of a combination bubble/hairpin, as
shown in Fig.  \ref{fig:fig2}B, which has a total of six variations. 
Note that the presence of the bubble permits the sharp kink between the
two connected duplex sections required at a vertex of the
dodecahedron.  There are now four RNA strands located at the vertex so
the electrostatic energy, $\Delta E_{B}$, of a class B vertex is
presumably higher than that of a class A vertex.  Finally, for a class
C vertex, three hair-pin/stem-loops are located at one vertex
A class C vertex requires that six RNA strands are located at a
single vertex and thus has presumably the highest electrostatic
energy ($\Delta E_{c}$).

The energy cost of a particular secondary RNA2 structure is, then,
\begin{equation}
H = E^{\rm sol} +N_{A} \Delta E_{A} + N_{B} \Delta E_{B} + N_{C} 
\Delta E_{C}
\label{ham}
\end{equation}
with $E^{\rm sol}$ the solution energy cost of the secondary structure
computed by the existing procedure \cite{bundhwa}, while $N_{A \!  -
\!  C}$ denotes the number of vertices.  The non-specific
electrostatic binding energy contributes a constant energy that need
not be included.

\begin{figure}[htb]
\includegraphics[height=1.2in]{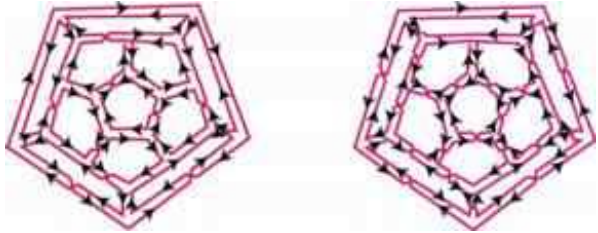}
\caption{Two examples of modified Euler paths covering the edges of
a dodecahedron.  The vertex at which each path begins and ends is in
the upper central portion of the dodecahedral graph.  The path at the
right contains the minimum number of crossings, while the path at the
right contains the maximum number of crossings.}
\label{fig:fig3}
\end{figure}

Since, by assumption, Class A vertices have the lowest free energy it
is logical to start by constructing RNA configurations that have
\emph{only} A-type junctions, in which case there are no hairpins or
stem-loops at all (i.e. $N_{B} = N_{C} =0$).  That does \textbf{not}
mean that the genome has to be disconnected.  Recall that there are
two different possible junctions, related by a $180^{\circ}$ rotation. 
Since we allow for vertex heterogeneity, we can use both junction
types and search for a single-connected genome.  In order to do this,
we represented the dodecahedral cage by a planar graph.  A
singly-connected RNA molecule must visit every edge of this graph
twice, each case in opposite directions, in order to form the double
helical segments on the edges.  \emph{The existence of this type of
path is not forbidden by graph theory}.  The problem of identifying a
permissible path for a ss RNA molecule is equivalent to a finding
particular choice for the A1 and A2 vertices at every node of the
dodecahedron such that they can be connected by a continuous path.  We
will call this a \emph{Modified Euler Path} (MEP).  In order to
identify MEP's, we carried out a computer search in which one of the
nodes of the dodecahedron was singled out as the starting (and ending)
point of the MEP, while all other vertices are either designated as
right turn or left turn vertices.  This produced 262,144 different
candidates for a MEP. Most of the candidate walks were unsuccessful,
because the walker arrived at a point at which the conditions for the
walk were violated.  However, in 54,272 of the cases a MEP was
generated with a mean success rate of 53/256=0.207\ldots Two examples
of such walks are displayed in Fig.  \ref{fig:fig3}.
Although the (3D) symmetry of these RNA structures is very close
to icosahedral, it is not exact because the successful MEPÕs contain a
certain number of edges with an extra crossing, as indicated by a
cross in Fig.  \ref{fig:fig3}.  This implies that the double-helical
segments that occupy these must have either an extra half turn, or a
deficit of one half turn.  The MEPÕs we generated contained a minimum
of eleven crossings and a maximum of twenty, corresponding to the two
cases shown in Fig.  \ref{fig:fig3}.  If the energy cost of Type B and
Type C vertices are significantly higher than that of a Type A vertex,
then the pattern with the minimum number of extra crossings should be the
lowest free energy structure.  

However, it is well known that a path covering a graph with overpasses
and underpasses, as shown in Fig.  \ref{fig:fig3}, are \emph{knotted}
\cite{knots}.  Even the MEP that has the minimum number of crossings
still is highly knotted.  RNA molecules in solution are never knotted
(though viral RNA may have ``pseudo-knots'' \cite{pseudo}), and a
knotted RNA molecule released by a virus during infection would not be
functional.  We thus must exclude knotted RNA structure and hence
Class A Quasi-Icosahedral order.
Knot-free genomes can be constructed by demanding that, unlike the
structures shown in Fig.  \ref{fig:fig3}, the secondary structure of
RNA molecules inside a virus has the same linearly-branched,
circuit-free topology as RNA molecules in solution.  When a linearly
branched structure decorates a dodecahedral graph, types B and C
vertices unavoidably appear.  Given the presumed higher free energy
cost of a type C vertex, we will construct linearly branched genomes
with only type A and B vertices.  Linearly-branched RNA secondary
structures characteristically contains an equal number of branches and
hairpin/stemloop structures.  That means that \emph{there must be an
equal number of A and B type vertices}, so $N_{A}=N_{B}$.  The
simplest example of such an A-B structure involves singly-branched RNA
molecules, i.e. the molecule consists of a \emph{main-chain} and a
certain number of unbranched \emph{side-chains}.  Secondary structures
that are nearly singly-branched are indeed encountered in the spectrum
of RNA2, with an energy of about 10 kcal/mole above the groundstate
\cite{pseudo}.  

\begin{figure}[htb]
\includegraphics[height=2in]{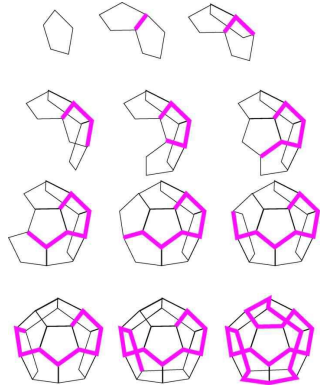}
\caption{Assembly of a viral capsid governed by coordinated with the 
tracing out of a Hamiltonian path by the icosahedrally-ordered RNA. 
For the fully-assembled capsid, shown in the lower right-hand corner 
of the figure, all portions of the Hamiltonian path traced out by the 
RNA main chain are indicated.}
\label{fig:fig4}
\end{figure}

To cover the dodecahedral graph with a single-branched structure,
every edge of the dodecahedron must be a link between an A type and a
B type vertex.  If this is the case, then the main chain of the RNA
molecule must visit every vertex of the dodecahedral graph {\em once
and only once}.  The construction of such a route is a well-known
problem in graph theory, known as a ``Hamiltonian Path''
\cite{bollobas,note}.  We will focus on the special case of a
Hamiltonian Path that starts and ends on the same vertex of a graph,
which is a Hamiltonian Cycle, is shown in the lower right-hand corner
of Fig.  \ref{fig:fig4}.  In this case, any point on the graph can be
treated as the starting site.  To every edge of the graph that is not
part of the Hamiltonian Path, we must assign an A-B or a B-A pair of
vertices, so for the Hamiltonian Cycle of relevance to the ordering of
RNA on an icosahedral virus, we obtain $2^{10}$ different possible
configurations.  These $2^{10}$ configurations form a subset of the
allowed secondary RNA stuctures in solution.  For a given base-pair
sequence, the optimal secondary structure within this subset can, in
principle, be determined by minimizing $H$.

The biological relevance of the secondary structure of the viral
genome is, in fact, intimately connected with the {\em assembly
scenario} of the virus, and this imposes important constraints on the
RNA configuration.  It has been well-established for ss RNA
cylindrical viruses that the RNA molecule acts as a linear growth
template \cite{virology}.  Growth starts from a specific hairpin
structure (the ``packaging signal'') and proceeds by successive
addition of protein oligomers under the action of the non-specific
protein-RNA interaction.  The length of the RNA molecule determines
the size of the virus.  It is currently not known whether main chain
type ss-RNA molecules also can act as growth templates for the
icosahedral family of viruses, but the only microscopy study of the
growth kinetics of RNA/capsid co-assembly of an icosahedral virus does
report that growth starts from a three-protein nucleus followed by the
successive addition of oligomer units \cite{sorger}.  The observed
co-assembly scenario resmbles the growth of a curved two-dimensional
(2D) \emph{protein crystal} that closes in on itself.  The
characteristic feature of conventional (slowly) growing crystals is
that they are compact.  New units are predominantly added to those
available sites on the surface of the crystal where the new unit can
make a maximum number of bonding contacts with units that are already
incorporated.  The co-assembly condition requires the presence of RNA
material when new units are added.  The effect of this process is to
even out surface roughness, although with increased growth velocities,
the growth surface may roughen.

To examine whether the reported growth scenario is consistent with the
proposed Hamiltonian path/cycle model developed above, we assume that
the Hamitonian Path construction provides the lowest accessible free
energy structure as proposed in the previous section.  The assembly
history for the Hamiltonian Cycle is shown in Fig.  \ref{fig:fig4}, in
which we show only the main chain part.  Edges that are not indicated
by heavy lines in the figure below are occupied by a side-chain,
ending in a hairpin.  Except for the first step, every added pentamer
bonds at least to two edges, one occupied by the main chain and one
occupied by a side chain.  The growth morphology imposed by the
Hamiltonian path construction is, then, consistent with compact 2D
growth structures.

In summary, on the basis of elementary graph-theoretical arguments, we
propose that the quasi-icosahedral organization of unsegmented ss RNA
viral genomes is based on the exploitation of the secondary structure
of the viral RNA to form a Hamiltonian path or cycle along a subset of
the edges of a polyhedron inscribed on the capsid surface, with
side-branches covering the remaining edges.  The important role for
the RNA main-chain structure would be consistent with the main-chain
acting as a linear template for the growth of the capsid during the
self-assembly process.  Our arguments predict that quasi-icosahedral
organization should never be encountered for viruses with unsegmented
fully duplexed ds DNA or RNA genomes.  Experimental tests should be
straightforward.  For instance, FHV self-assembles under in-vitro
conditions.  Self-assembly of FHV with a duplexed RNA strand having
the same length as the FHV ss RNA should be difficult or impossible.  
Determining the nature fo the vertex inhomogeneity of actual 
Nodaviridae would provide a decisive test of the model. Because 
current structural determination methods {\em impose} icosahedral 
symmetry, this is not yet possible, but we are hopeful that 
developments in these methods will make such a test feasible in the 
not-too-distant future.

We would like to thank Professors W. M. Gelbart and J. E. Johnson for 
helpful discussions. We are also grateful to Professor Johnson for 
permission to reproduce Fig. \ref{fig:fig1}.

\bibliography{euler}

\begin{thebibliography}{21}
\expandafter\ifx\csname natexlab\endcsname\relax\def\natexlab#1{#1}\fi
\expandafter\ifx\csname bibnamefont\endcsname\relax
  \def\bibnamefont#1{#1}\fi
\expandafter\ifx\csname bibfnamefont\endcsname\relax
  \def\bibfnamefont#1{#1}\fi
\expandafter\ifx\csname citenamefont\endcsname\relax
  \def\citenamefont#1{#1}\fi
\expandafter\ifx\csname url\endcsname\relax
  \def\url#1{\texttt{#1}}\fi
\expandafter\ifx\csname urlprefix\endcsname\relax\def\urlprefix{URL }\fi
\providecommand{\bibinfo}[2]{#2}
\providecommand{\eprint}[2][]{\url{#2}}

\bibitem[{\citenamefont{Bink and Pleij}(2002)}]{bink}
\bibinfo{author}{\bibfnamefont{H.~H.} \bibnamefont{Bink}} \bibnamefont{and}
  \bibinfo{author}{\bibfnamefont{C.~W.} \bibnamefont{Pleij}},
  \bibinfo{journal}{Arch Virol} \textbf{\bibinfo{volume}{147}},
  \bibinfo{pages}{2261} (\bibinfo{year}{2002}).

\bibitem[{\citenamefont{Odijk}(1998)}]{odijk}
\bibinfo{author}{\bibfnamefont{T.}~\bibnamefont{Odijk}},
  \bibinfo{journal}{Biophysical Journal} \textbf{\bibinfo{volume}{75}},
  \bibinfo{pages}{1223} (\bibinfo{year}{1998}).

\bibitem[{\citenamefont{Riemer and Bloomfield}(1978)}]{reimer}
\bibinfo{author}{\bibfnamefont{S.}~\bibnamefont{Riemer}} \bibnamefont{and}
  \bibinfo{author}{\bibfnamefont{V.~A.} \bibnamefont{Bloomfield}},
  \bibinfo{journal}{Biopolymers} \textbf{\bibinfo{volume}{17}},
  \bibinfo{pages}{785} (\bibinfo{year}{1978}).

\bibitem[{\citenamefont{Hud}(1995)}]{hud}
\bibinfo{author}{\bibfnamefont{N.~V.} \bibnamefont{Hud}},
  \bibinfo{journal}{Biophysical Journal} \textbf{\bibinfo{volume}{69}},
  \bibinfo{pages}{1355} (\bibinfo{year}{1995}).

\bibitem[{\citenamefont{Bundschuh and Hwa}(2002)}]{bundhwa}
\bibinfo{author}{\bibfnamefont{R.}~\bibnamefont{Bundschuh}} \bibnamefont{and}
  \bibinfo{author}{\bibfnamefont{T.}~\bibnamefont{Hwa}},
  \bibinfo{journal}{Phys. Rev. E.} \textbf{\bibinfo{volume}{65}},
  \bibinfo{pages}{031903/1} (\bibinfo{year}{2002}).

\bibitem[{\citenamefont{Fisher and Johnson}(1993)}]{fisher}
\bibinfo{author}{\bibfnamefont{A.~J.} \bibnamefont{Fisher}} \bibnamefont{and}
  \bibinfo{author}{\bibfnamefont{J.~E.} \bibnamefont{Johnson}},
  \bibinfo{journal}{Nature} \textbf{\bibinfo{volume}{361}},
  \bibinfo{pages}{176} (\bibinfo{year}{1993}).

\bibitem[{\citenamefont{Baker et~al.}(1999)\citenamefont{Baker, Olson, and
  Fuller}}]{3d}
\bibinfo{author}{\bibfnamefont{T.~S.} \bibnamefont{Baker}},
  \bibinfo{author}{\bibfnamefont{N.~H.} \bibnamefont{Olson}}, \bibnamefont{and}
  \bibinfo{author}{\bibfnamefont{S.~D.} \bibnamefont{Fuller}},
  \bibinfo{journal}{Microbiol Mol Biol Rev} \textbf{\bibinfo{volume}{63}},
  \bibinfo{pages}{862} (\bibinfo{year}{1999}).

\bibitem[{\citenamefont{Speir et~al.}(1995)\citenamefont{Speir, Munshi, Wang,
  Baker, and Johnson}}]{ccmv}
\bibinfo{author}{\bibfnamefont{J.~A.} \bibnamefont{Speir}},
  \bibinfo{author}{\bibfnamefont{S.}~\bibnamefont{Munshi}},
  \bibinfo{author}{\bibfnamefont{G.}~\bibnamefont{Wang}},
  \bibinfo{author}{\bibfnamefont{T.~S.} \bibnamefont{Baker}}, \bibnamefont{and}
  \bibinfo{author}{\bibfnamefont{J.~E.} \bibnamefont{Johnson}},
  \bibinfo{journal}{Structure} \textbf{\bibinfo{volume}{3}},
  \bibinfo{pages}{63} (\bibinfo{year}{1995}).

\bibitem[{\citenamefont{Tsuruta et~al.}(1998)\citenamefont{Tsuruta, Reddy,
  Wikoff, and Johnson}}]{fhv}
\bibinfo{author}{\bibfnamefont{H.}~\bibnamefont{Tsuruta}},
  \bibinfo{author}{\bibfnamefont{V.~S.} \bibnamefont{Reddy}},
  \bibinfo{author}{\bibfnamefont{W.~R.} \bibnamefont{Wikoff}},
  \bibnamefont{and} \bibinfo{author}{\bibfnamefont{J.~E.}
  \bibnamefont{Johnson}}, \bibinfo{journal}{J Mol Biol}
  \textbf{\bibinfo{volume}{284}}, \bibinfo{pages}{1439} (\bibinfo{year}{1998}).

\bibitem[{\citenamefont{Prasad et~al.}(1994)\citenamefont{Prasad, Matson, and
  Smith}}]{calicivirus}
\bibinfo{author}{\bibfnamefont{B.~V.} \bibnamefont{Prasad}},
  \bibinfo{author}{\bibfnamefont{D.~O.} \bibnamefont{Matson}},
  \bibnamefont{and} \bibinfo{author}{\bibfnamefont{A.~W.} \bibnamefont{Smith}},
  \bibinfo{journal}{J Mol Biol} \textbf{\bibinfo{volume}{240}},
  \bibinfo{pages}{256} (\bibinfo{year}{1994}), \bibinfo{note}{94300601
  0022-2836 Journal Article}.

\bibitem[{\citenamefont{Baker et~al.}(1992)\citenamefont{Baker, Cheng, Johnson,
  Olson, Wang, and Schmidt}}]{cmv}
\bibinfo{author}{\bibfnamefont{T.~S.} \bibnamefont{Baker}},
  \bibinfo{author}{\bibfnamefont{R.~H.} \bibnamefont{Cheng}},
  \bibinfo{author}{\bibfnamefont{J.~E.} \bibnamefont{Johnson}},
  \bibinfo{author}{\bibfnamefont{N.~H.} \bibnamefont{Olson}},
  \bibinfo{author}{\bibfnamefont{G.~J.} \bibnamefont{Wang}}, \bibnamefont{and}
  \bibinfo{author}{\bibfnamefont{T.~S.} \bibnamefont{Schmidt}}, in
  \emph{\bibinfo{booktitle}{Proc. Electron Microsc. Soc. Am.}}
  (\bibinfo{publisher}{San Francisco Press}, \bibinfo{address}{San Francisco},
  \bibinfo{year}{1992}), vol.~\bibinfo{volume}{50}, pp.
  \bibinfo{pages}{454--455}.

\bibitem[{\citenamefont{B\"{o}ttcher and Crowther}(1996)}]{tymv}
\bibinfo{author}{\bibfnamefont{B.}~\bibnamefont{B\"{o}ttcher}}
  \bibnamefont{and} \bibinfo{author}{\bibfnamefont{R.~A.}
  \bibnamefont{Crowther}}, \bibinfo{journal}{Structure}
  \textbf{\bibinfo{volume}{4}}, \bibinfo{pages}{387} (\bibinfo{year}{1996}).

\bibitem[{\citenamefont{Tang et~al.}(2001)\citenamefont{Tang, Johnson, Ball,
  Lin, Yeager, and Johnson}}]{tang}
\bibinfo{author}{\bibfnamefont{L.}~\bibnamefont{Tang}},
  \bibinfo{author}{\bibfnamefont{K.~N.} \bibnamefont{Johnson}},
  \bibinfo{author}{\bibfnamefont{L.~A.} \bibnamefont{Ball}},
  \bibinfo{author}{\bibfnamefont{T.}~\bibnamefont{Lin}},
  \bibinfo{author}{\bibfnamefont{M.}~\bibnamefont{Yeager}}, \bibnamefont{and}
  \bibinfo{author}{\bibfnamefont{J.~E.} \bibnamefont{Johnson}},
  \bibinfo{journal}{Nature Structural Biology} \textbf{\bibinfo{volume}{8}},
  \bibinfo{pages}{77} (\bibinfo{year}{2001}).

\bibitem[{tan()}]{tangnote}
\bibinfo{note}{Tang, et all \cite{tang} suggest that the paradox might be
  resolved once we allow the RNA1 molecule to ``mix'' in with RNA2 as part of
  the dodecahedral cage. Although possible, this is less elegant, in view of
  the dominant role that RNA2 plays in viral self-assembly.}

\bibitem[{\citenamefont{Bollob\'{a}s}(1990)}]{bollobas}
\bibinfo{author}{\bibfnamefont{B.}~\bibnamefont{Bollob\'{a}s}},
  \emph{\bibinfo{title}{Graph theory: an introductory course}}, Graduate texts
  in mathematics ; 63. (\bibinfo{publisher}{Springer Verlag},
  \bibinfo{address}{New York}, \bibinfo{year}{1990}),
  \bibinfo{edition}{corrected third printing.} ed.

\bibitem[{gra()}]{graphnote}
\bibinfo{note}{The actual theorem states that an Euler path is impossible if
  there are more than two vertices connected to an odd number of edges. In both
  the dodecahedron and the icosahedron all vertices have an odd number of edges
  incident.}

\bibitem[{\citenamefont{Nechaev}(1996)}]{knots}
\bibinfo{author}{\bibfnamefont{S.~K.} \bibnamefont{Nechaev}},
  \emph{\bibinfo{title}{Statistics of knots and entangled random walks}}
  (\bibinfo{publisher}{World Scientific}, \bibinfo{address}{Singapore ; River
  Edge, NJ}, \bibinfo{year}{1996}), \bibinfo{note}{s.K. Nechaev. Includes
  bibliographical references and index.}

\bibitem[{\citenamefont{Pleij et~al.}(1985)\citenamefont{Pleij, Rietveld, and
  Bosch}}]{pseudo}
\bibinfo{author}{\bibfnamefont{C.~W.} \bibnamefont{Pleij}},
  \bibinfo{author}{\bibfnamefont{K.}~\bibnamefont{Rietveld}}, \bibnamefont{and}
  \bibinfo{author}{\bibfnamefont{L.}~\bibnamefont{Bosch}},
  \bibinfo{journal}{Nucleic Acids Res} \textbf{\bibinfo{volume}{13}},
  \bibinfo{pages}{1717} (\bibinfo{year}{1985}).

\bibitem[{not()}]{note}
\bibinfo{note}{An interesting historical footnote is that the notion of a
  Hamiltonian path originated in the construction by William Rowan Hamilton of
  a puzzle, called the ``Icosian Game,'' the object of which is to find a
  Hamiltonian path along the vertices of a dodecahedron.}

\bibitem[{\citenamefont{Flint}(2000)}]{virology}
\bibinfo{author}{\bibfnamefont{S.~J.} \bibnamefont{Flint}},
  \emph{\bibinfo{title}{Principles of virology : molecular biology,
  pathogenesis, and control}} (\bibinfo{publisher}{ASM Press},
  \bibinfo{address}{Washington, D.C.}, \bibinfo{year}{2000}).

\bibitem[{\citenamefont{Sorger et~al.}(1986)\citenamefont{Sorger, Stockley, and
  Harrison}}]{sorger}
\bibinfo{author}{\bibfnamefont{P.~K.} \bibnamefont{Sorger}},
  \bibinfo{author}{\bibfnamefont{P.~G.} \bibnamefont{Stockley}},
  \bibnamefont{and} \bibinfo{author}{\bibfnamefont{S.~C.}
  \bibnamefont{Harrison}}, \bibinfo{journal}{J Mol Biol}
  \textbf{\bibinfo{volume}{191}}, \bibinfo{pages}{639} (\bibinfo{year}{1986}).

\end{thebibliography}

\pagebreak

\end{document}